\documentclass[twocolumn,aps,prl,superscriptaddress]{revtex4-1}

\usepackage{hyperref}
\usepackage{graphicx}
\usepackage{amsmath}
\usepackage{amssymb}
\usepackage{xcolor}
\usepackage{bm}
\usepackage{color}
\usepackage{enumitem} 
\usepackage{lipsum}
\usepackage{amsmath, array} 
\usepackage[normalem]{ulem}
\usepackage{soul,xcolor}
\usepackage{lineno}

\usepackage[export]{adjustbox}
\usepackage{subfig}
\usepackage{floatrow}

\begin{document}
\title{Photo-induced two-body loss of ultracold molecules} 

\author{Arthur Christianen} 
\affiliation{Institute for Molecules and Materials, Radboud University, Nijmegen, The Netherlands}

\author{Martin W. Zwierlein}
\affiliation{MIT-Harvard Center for Ultracold Atoms, Research Laboratory of Electronics, and Department of Physics, Massachusetts Institute of Technology, Cambridge, Massachusetts 02139, USA}

\author{Gerrit C. Groenenboom} 
\affiliation{Institute for Molecules and Materials, Radboud University, Nijmegen, The Netherlands}

\author{Tijs Karman}
\email{tijs.karman@cfa.harvard.edu}
\affiliation{\!ITAMP, \!Harvard-Smithsonian \!Center \!for \!Astrophysics, \! \!Cambridge, \!Massachusetts, \!02138, \!USA}

\begin{abstract}
The lifetime of nonreactive ultracold bialkali gases was conjectured to be
limited by sticky collisions amplifying three-body loss. We show that the
sticking times were previously overestimated and do not support this
hypothesis. We find that electronic excitation of NaK+NaK collision
complexes by the trapping laser leads to the experimentally observed
two-body loss. We calculate the excitation rate with a quasiclassical,
statistical model employing \emph{ab initio} potentials and transition
dipole moments. Using longer laser wavelengths or repulsive box potentials
may suppress the losses.
\end{abstract}

\maketitle

Ultracold dipolar gases have exciting applications across physics and
chemistry \cite{carr:2009,krems:09}. They can be used for high-precision
measurements that challenge the Standard Model of particle physics
\cite{baron:2014,andreev:2018}, to model or simulate quantum many-body
physics \cite{micheli:2006,buchler:2007,cooper:2009}, and to study and
control chemical reactions \cite{krems:08, ospelkaus:2010b}. There are also
promising schemes for quantum computation using ultracold dipolar gases
\cite{demille:2002, yelin:2006,ni:2018}.

The first ultracold dipolar molecules in their absolute rovibronic and
hyperfine ground state were realized by Ospelkaus \emph{et al.}
\cite{ospelkaus:2010a}.  The KRb molecules used were reactive
\cite{ni:2008,ospelkaus:2010b}, which limits the lifetime of these ultracold
gases. To avoid losses due to chemical reactions, several groups realized
nonreactive ultracold dipolar gases of the bosonic $^{87}$Rb$^{133}$Cs
\cite{takekoshi:2014, molony:2014}, $^{23}$Na$^{87}$Rb \cite{guo:2016} and
the fermionic $^{23}$Na$^{40}$K \cite{park:2015,seesselberg:2018}. However,
losses were still observed at about the same rate that would be expected of
reactive molecules \cite{ye:2018,guo:18,gregory:19}. In all of these
experiments the lifetime of the gas in the crossed optical dipole trap was
limited to a few seconds \cite{park:2015} or less \cite{takekoshi:2014,
molony:2014,guo:2016}.  Preventing this loss is of crucial importance for
realizing higher molecular densities required for loading optical lattices
and to improve the coherence time of ultracold molecules \cite{park:2017}.

The loss mechanism in these nonreactive gases is not yet understood, but
there are strong indications that the loss is caused by ultracold molecular
collisions, which have been studied extensively in the literature
\cite{quemener:2012}. Mayle \emph{et al.}\ \cite{mayle:2012,mayle:2013}
proposed that the loss mechanism may be due to ``sticky collisions'', the
formation of long-lived collision complexes. What actually happens to these
complexes that causes loss of molecules from the trap is unknown and the
subject of this paper. We show that laser excitation of these complexes can
explain the losses in the experiments.

Mayle \emph{et al.}\ \cite{mayle:2012,mayle:2013}  propose a procedure to
calculate the sticking time of a collision complex by calculating the
density of states. Rice-Ramsperger-Kassel-Marcus (RRKM) theory
\cite{levine:2005} relates the sticking time $\tau_\mathrm{stick}$ of the
collision to the density of states (DOS, $\rho$)
\begin{equation}
  \tau_\mathrm{stick}=2 \pi \hbar \rho.
\end{equation}
Mayle \emph{et al}. obtain the RRKM sticking time using multichannel  quantum defect theory (MQDT) [\cite{mayle:2012,mayle:2013},
which treats the long range fully quantum mechanically but uses a simplified short range, parameterized by the DOS. However, in the accompanying paper \cite{christianen:2019b} we show there was an error in the DOS calculation, leading to sticking times of two to three orders of magnitude too large. 
A quasiclassical equation
to accurately calculate the DOS for arbitrary potentials is found:
\begin{equation} \label{eq:DOScalc}
  \rho=\frac{g_{\bm{N}Jp}8 \pi^{3+\frac{D}{2}} \hbar^3
       C_{\bm{Nm}} (2J+1)}{\Gamma(\frac{D}{2})} \int\! G(\bm{q})
  [E-V(\bm{q})]^{\frac{D}{2}-1} d\bm{q}.
\end{equation}
Here, $E$ is the total energy in the system, $J$ is the total angular
momentum quantum number, $p$ is the parity and $V(\bm{q})$ is the potential
energy as a function of the Jacobi coordinates of the complex
$\bm{q}=(R,r_1,r_2,\theta_1, \theta_2,\phi)$: the intermolecular distance
$R$, the bond lengths $r_{1}$, $r_2$, the polar angles, $\theta_{1}$,
$\theta_2$, and the dihedral angle, $\phi$ 
(see Ref.~\cite{christianen:2019}). The
complex has $D=6$ internal degrees of freedom.  The constant $C_{\bm{Nm}}$,
parity factor $g_{\bm{N}Jp}$,  and geometry factor $G(\bm{q})$ are defined
in the accompanying paper \cite{christianen:2019b}.  The factor
$g_{\bm{N}Jp}=1/2$ for collisions of heteronuclear diatoms. 

We use Eq.~\eqref{eq:DOScalc} and the recently constructed, accurate
potential energy surface (PES) \cite{christianen:2019} to calculate the DOS
for the NaK+NaK system. These calculations result in a DOS of
$0.37~\mu$K$^{-1}$ (for $J=1$) and a sticking time of about 18~$\mu$s.  We
showed that this lifetime is not long enough for complex-diatom collisions
to explain the experimental losses \cite{christianen:2019b}.

\begin{figure}[!t]
\centering
\includegraphics[height=2.4in]{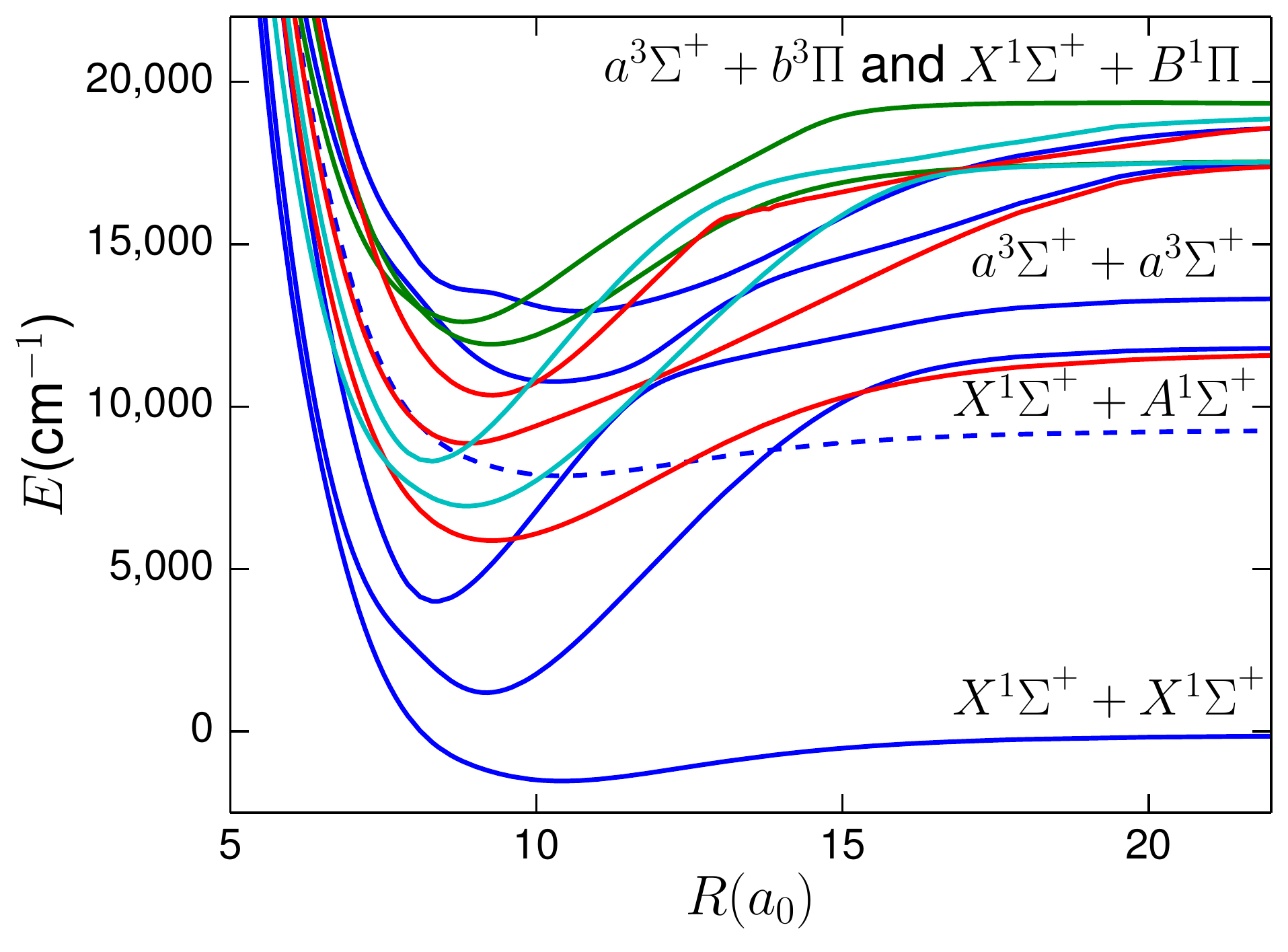}
\captionsetup{format=plain,justification=RaggedRight,singlelinecheck=false}
\caption{A one-dimensional cut of the NaK+NaK PESs for the ground state and
several excited states. The geometry is $r_1=r_2=6.9$~$a_0$,
$\theta_1=3\pi/4$, $\theta_2=\pi/4$, $\phi=\pi$. \emph{Ab initio} points
were calculated over the entire range of $R$ with intervals of
$0.5~a_0$. The dashed line indicates the ground state energy curve plus the
photon energy $\hbar \omega$ of a 1064 nm laser. With this geometry, the
complex has a $C_{2h}$ symmetry. The colors in the graph correspond to
the irreducible representations of the states, blue
for $A_g$, green for $A_u$, red for $B_u$, and cyan for $B_g$.}
\label{fig:exstates}
\end{figure}

In this paper, we show that the excitation of collision complexes by the
trapping laser can explain the losses observed experimentally.  In typical
experiments, the diatoms are confined using a crossed optical dipole trap
with lasers far red detuned (1064~nm \cite{takekoshi:2014,ye:2018,
park:2015,park:2017} or 1550 nm
\cite{molony:2014,seesselberg:2018,seesselberg:2018b}) from the molecular
$X^1\Sigma^+ \rightarrow A^1\Sigma^+$ transition.  However, the electronic
excitation energies of the complex differ from those of the individual
molecules and depend on the nuclear geometry of the complex.  This means
that even though the laser is red detuned for the diatoms, the laser may
electronically excite the collision complex.

Figure~\ref{fig:exstates} shows the ground state and low-lying singlet
excited states of NaK+NaK as a function of the intermolecular distance for a
planar configuration of $C_{2h}$ symmetry.  Colors correspond to the
different irreducible representations of the electronic states.  The
potentials were calculated \emph{ab initio} using internally contracted
multireference configuration interaction with {\sc Molpro}
\cite{molpro:2015}, as described in more detail in the Supplement
\cite{supplement}.  The lowest two excited states correlate asymptotically 
to $X^1\Sigma^+ \rightarrow A^1\Sigma^+$ excitations of one of the NaK
molecules.  The third excited state of the complex correlates to both
molecules excited to $a ^3\Sigma^+$.  Above that, we find excited states
correlating to single $X^1\Sigma^+ \rightarrow B^1\Pi$ excitations, as well
as simultaneous excitation of one molecule to $a^3\Sigma^+$ and the other
to $b {^3\Pi}$. These highest two thresholds are nearly degenerate and their
order depends on the monomer bond lengths.  Higher excited states
correlating to Na$(S\rightarrow P)$ transitions and more highly excited
triplet states exist, but do not contribute to the absorption.  We see that
the excited-state potentials are more strongly bound than the ground-state
potential, e.g., the first excited state potential has a well depth in the
order of 12\,000 cm$^{-1}$ compared to the well depth of 4534 cm$^{-1}$ of the
ground-state potential \cite{christianen:2019}.  The dashed line shows the
ground-state potential shifted up by 1064~nm.  This curve crosses four
excited-state potentials in the region where the ground-state potential is
attractive, indicating that these excited states can be reached.
Furthermore, strong transition dipole moments (TDMs) to these excited states
exists as the NaK $A^1\Sigma^+$ and $B^1\Pi$ excited states correlate to
parallel and perpendicular dipole-allowed K$(S\rightarrow P)$ transitions.

This establishes that the energies of the excited states are low enough for
trapping-laser-induced electronic transitions of collision complexes to
occur.  Next, we calculate the rates of these transitions to determine
whether these transitions occur within the lifetime of a complex. This is not possible in the current MQDT framework, since this model does not explicitly treat the short range. Therefore, we develop a statistical model to calculate the laser excitation rates using \emph{ab initio} short-range potentials.

The rate equation for laser excitation of the complex for a frozen geometry,
$\bm{q}$, from discrete state $\mathrm{i}$ to $\mathrm{f}$ is given
by\cite{hilborn:82} \begin{equation} \label{eq:exrate} W_{\mathrm{i
\rightarrow f}}(\bm{q})=- \int d\omega \frac{b_{\mathrm{i\rightarrow
f}}(\bm{q},\omega) n}{c} \frac{dI}{d\omega}, \end{equation} where $n$ is the
particle density and $dI/d\omega$ is the spectral irradiance of the
laser. The coefficient $b_{\mathrm{i \rightarrow f}}$ is given by
\cite{hilborn:82}
\begin{equation}
  b_{\mathrm{i \rightarrow f}} (\bm{q},\omega)=\frac{\pi}{3 \epsilon_0
  \hbar^2} \mu_\mathrm{i \rightarrow f}(\bm{q})^2 g(\bm{q},\omega),
\end{equation}
where $\mu_{\mathrm{i\rightarrow f}}(\bm{q})$ is the TDM and
$g(\bm{q},\omega)$ is the lineshape of the transition.  We set $n=n_c$, the
number of collision complexes. We assume the collisions to be
ergodic and use the same quasiclassical phase-space model
used to calculate the sticking time.  The nuclear motion is treated
classically and electronic transitions can only occur when the ground-state
electronic energy plus the photon energy matches the excited-state
electronic energy, i.e., where the excited potentials cross the blue dashed
line in Fig.~\ref{fig:exstates}.

We calculate the expectation value of $b_{\mathrm{i \rightarrow f}} ({\bm
q},\omega)$ over the accessible phase space as a function of $\omega$. We
assume the linewidth of the transition is small compared to
the variation of the energy gap between the ground state and the excited
state with geometry, $E_{\mathrm{gap}}(\bm{q})$, and replace
$g(\bm{q},\omega)$ by a delta function,
\begin{multline} \label{eq:laserintegral}
  \langle b_{\mathrm{i \rightarrow f}}({\bm q},\omega) \rangle_{\bm q} =
  \frac{g_{\bm{N}Jp} 8 \pi^{3+\frac{D}{2}} \hbar^3
  C_{\bm{Nm}} (2J+1)}{3 \epsilon_0 \hbar^2 \Gamma(\frac{D}{2}) \rho } \\
  \int\!
  d \bm{q}\, G(\bm{q}) [E_\mathrm{tot}-V(\mathbf{y})]^{\frac{D}{2}-1}
  \mu_\mathrm{i \rightarrow f}^2(\bm{q})
  \delta\left[\frac{E_\mathrm{gap}(\bm{q})}{\hbar}-\omega\right].
\end{multline}
The linewidth of the laser is also very small with respect to the variation
of the excitation energy as a function of the geometry, such that
$dI/d\omega$ can be replaced by ${I_\mathrm{tot}
\delta(\omega_\mathrm{laser}-\omega)}$.  We then sum over the electronic
final states (f) to obtain
\begin{equation} \label{eq:laserexrate}
  \frac{dn_c}{dt}=-n_c\sum_\mathrm{f} \Gamma_\mathrm{i
  \rightarrow f} =-n_c \sum_\mathrm{f} \frac{ I_\mathrm{tot} \langle
  b_{\mathrm{i \rightarrow f}}(\omega_\mathrm{laser})\rangle}{c},
\end{equation}
where $\Gamma_\mathrm{i \rightarrow f}$ is the excitation rate from state i
to f. We define $\Gamma_{\mathrm{laser}}=\sum_\mathrm{f} \Gamma_\mathrm{i
\rightarrow f}$ and $\tau_{\mathrm{laser}}=\Gamma_{\mathrm{laser}}^{-1}$.

\begin{figure}[!t]
\centering
\includegraphics[height=2.4in]{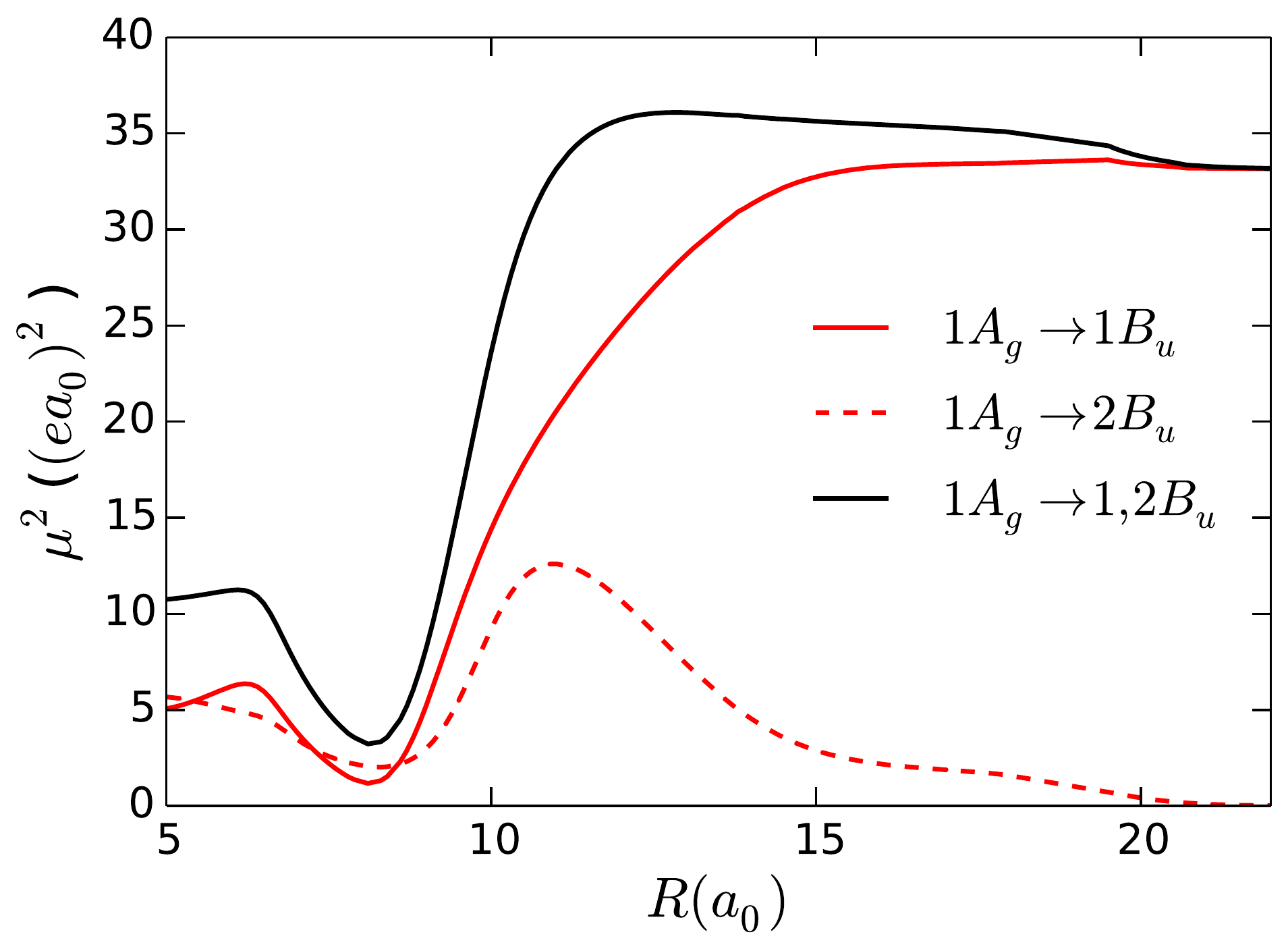}
\captionsetup{format=plain,justification=RaggedRight,singlelinecheck=false}
\caption{A one-dimensional cut of two NaK+NaK squared-TDM surfaces. The geometry is
$r_1=r_2=6.9~a_0$, $\theta_1=3\pi/4$, $\theta_2=\pi/4$, and $\phi=\pi$. The TDMs
plotted are those between the ground state and the first two excited states
with $B_u$ symmetry, corresponding to the red lines in Fig.\
\ref{fig:exstates}. The black line corresponds to the sum of those two
squared TDMs.}
\label{fig:exstates_tdm}
\end{figure}

Before evaluating Eq.~\eqref{eq:laserexrate}, we first qualitatively explore
the properties of the TDM between the ground state and the low-lying excited
states by again considering the one-dimensional cut for the $C_{2h}$ configuration.  Due
to symmetry, only $A_u$ and $B_u$ states have nonzero TDMs with the ground
state.  We show the TDMs of the two lower-lying $B_u$ states in
Fig.~\ref{fig:exstates_tdm}.  At long range, the $1B_{u}$
state corresponds to $(|A X\rangle - |X A\rangle)/\sqrt{2}$, the
anti-symmetric combination of $X^1\Sigma^+$ to $A^1\Sigma^+$ excitations in
either molecule.  The $A$ excited state correlates to K$(S \rightarrow P)$
transitions.  The molecular TDMs add constructively such that the $1B_{u}$
linestrength approaches twice the linestrength of the $S \rightarrow P$
transition of the K atom, $\mu^2 \sim 17$~$(ea_0)^2$.\ \cite{nandy:2012}.
Asymptotically, the $2B_u$ state corresponds to $a^3\Sigma^++b^3\Pi$, which
has zero TDM with the ground state.  For $R<20$~$a_0$, these states mix such
that the $1B_{u}$ TDM decreases and the $2B_{u}$ TDM increases, but the sum
remains nearly constant.  At yet shorter distances, $R<10$~$a_0$, short-range
effects decrease all TDMs.  The more highly excited states can be reached
energetically only at very short range, where the TDMs drop substantially,
such that the lowest three excited states dominate the excitation rate.  In
what follows, we include only these three excited states.

The evaluation of Eq.~\eqref{eq:laserintegral} requires global PESs and TDM
surfaces, rather than the one-dimensional cuts discussed above.  For the ground state, we
use the GP9 PES from Ref.~\cite{christianen:2019}.  New PESs for the lowest
three excited states are constructed to describe the gaps between the
electronic energy levels.  We also construct TDM surfaces for these excited
states, and fit all surfaces using our machine-learning fitting method
\cite{christianen:2019}.  The details are described in the Supplementary
Information \cite{supplement}. 

For geometries with lower symmetry, the excited states exhibit many avoided
crossings and conical intersections that complicate the electronic structure
calculations.  Conical intersections are expected to occur, e.g., from
the many crossings observed in Fig.~\ref{fig:exstates_tdm}, and because the
energetic ordering of electronic states switches between arrangements; the
$a^3\Sigma^++a^3\Sigma^+$ state is the third excited state in the NaK+NaK
arrangement, but the second excited state for Na$_2+$K$_2$.  Intersections
occur at intermediate geometries.  Near conical intersections, the TDM and
potential energy vary rapidly with nuclear geometry, and the individual
surfaces are difficult to fit accurately.  However, we are interested in the
laser-excitation rate, which is computed as a phase-space average summed
over electronically excited states, and is less sensitive to the quality of
the fits for individual electronic states.

The global TDM surfaces constructed here exhibit features similar to those
observed at the $C_{2h}$ symmetric configuration.  At long range, the
excited states have large TDMs due to their K$(S\rightarrow P)$ character.
For lower-symmetry configurations, the lowest three excited states mix
strongly and the linestrength distributes over the excited states, such that
the individual squared TDMs vary with geometry, but the total remains
approximately constant.  At yet shorter range, the total TDM decreases.

\floatsetup[figure]{style=plain,subcapbesideposition=top}
\begin{figure}[!t]
\centering
\includegraphics[height=2.4in]{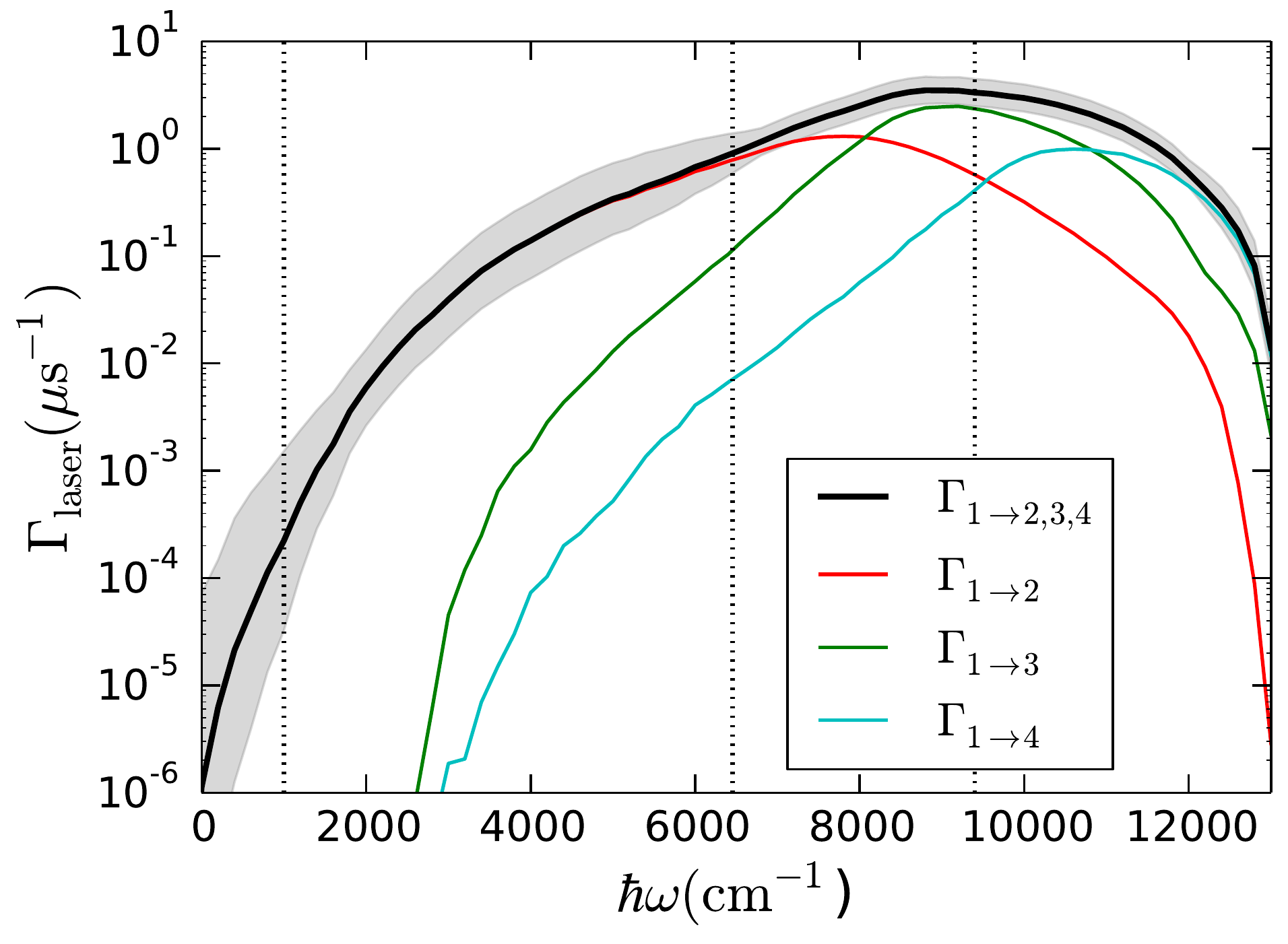}
\captionsetup{format=plain,justification=RaggedRight,singlelinecheck=false}
\caption{The calculated laser excitation rate to the first three excited
states as a function of the laser photon energy $\hbar \omega$. The vertical
black dotted lines indicate the energies of the 10 $\mu$m, 1550 nm, and 1064
nm lasers. To calculate $\Gamma_\mathrm{laser}$, a trap
depth of 10 $\mu$K was used. For the above wavelengths, this corresponds to
laser intensities of respectively 12.6, 10.2, and 7.3~kW~cm$^{-2}$. The
grey shaded region indicates the error margin, which is estimated as
described in the Supplement \cite{supplement}.}
\label{fig:rate_vs_freq}
\end{figure}

Figure~\ref{fig:rate_vs_freq} shows the laser excitation rate as a function
of the frequency at a trap depth of 10$\mu$K. The colored lines indicate the
laser transition rates to the individual excited states, the black line
indicates the total. The grey shaded area indicates the error margin, which
is estimated as described in the Supplement \cite{supplement}.  The vertical
dotted lines indicate wavelengths of 10~$\mu$m, 1550~nm, and 1064~nm.
At 1064 nm, the loss rate is $\Gamma_\mathrm{laser} = 3.3$~$\mu$s$^{-1}$,
which corresponds to a lifetime for laser excitation of
$\tau_\mathrm{laser}=0.30$~$\mu$s.  At 1550~nm, the loss rate is reduced to
0.91~$\mu$s$^{-1}$, corresponding to a lifetime of 1.1~$\mu$s.  At either
wavelength, the lifetime for laser excitation is much smaller than the
sticking time, $\tau_\mathrm{laser}\ll \tau_\mathrm{stick}=18$~$\mu$s.  Hence, essentially
all complexes formed undergo laser excitation before they dissociate, such
that complex formation manifests as effective two-body loss, in agreement
with experimental observations in
Refs.~\cite{takekoshi:2014,molony:2014,park:2015,ye:2018,gregory:19}.
Switching to a 10~$\mu$m laser wavelength would reduce the excitation rate
by orders of magnitude to around 0.2 ms$^{-1}$, which is much slower than
the complex dissociation rate.

\begin{figure}[!t]
\includegraphics[height=2.5in]{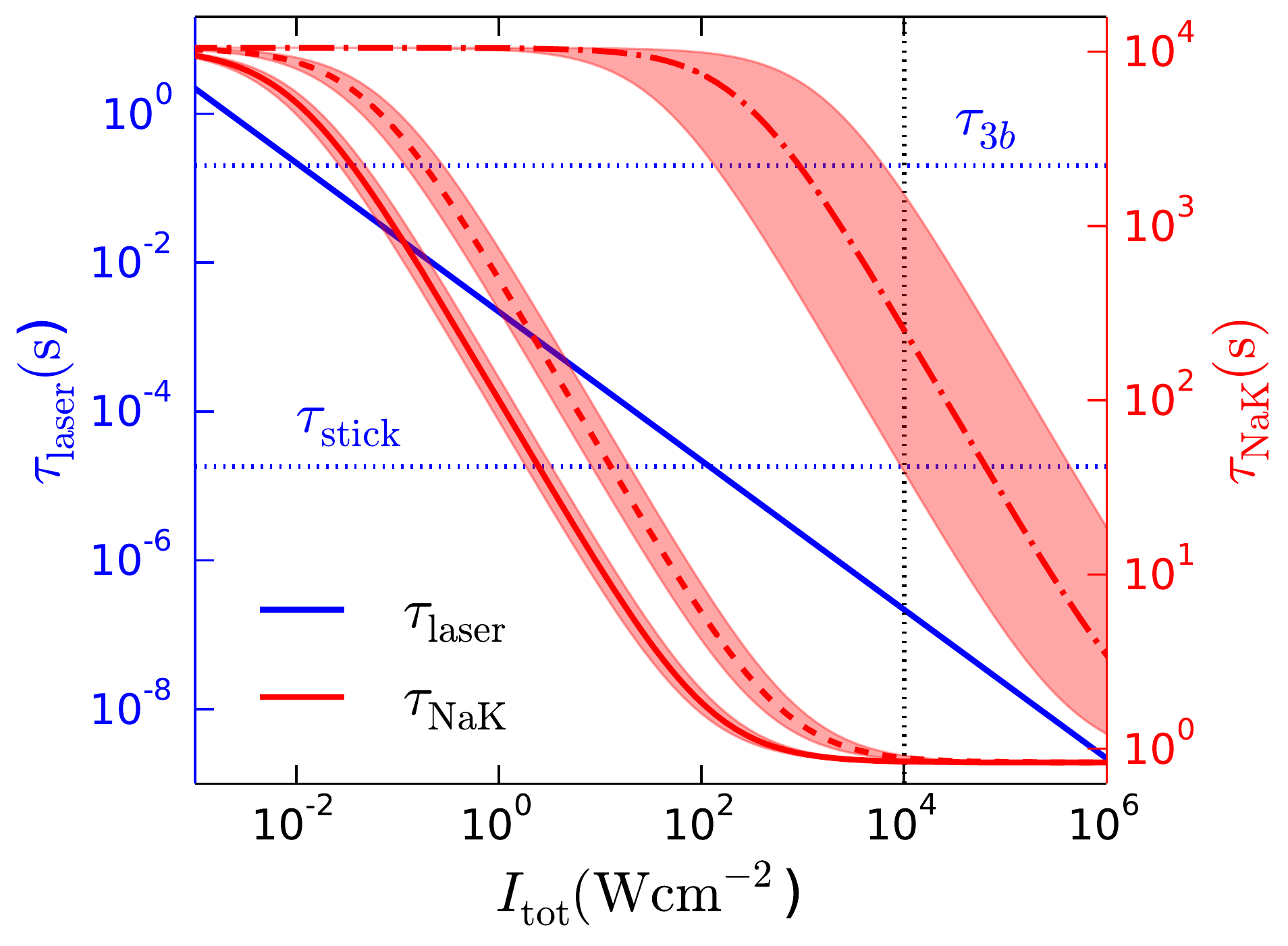}
\captionsetup{format=plain,justification=RaggedRight,singlelinecheck=false}
\caption{The laser-excitation lifetime, $\tau_\mathrm{laser}$ (in blue,
left-hand axis, for a 1064 nm laser) and the half-life of
NaK molecules in a crossed optical dipole trap, $\tau_\mathrm{NaK}$ (in red,
right-hand axis), as a function of the laser intensity.  The solid, dashed,
and dash-dotted red lines show results obtained using the full \emph{ab
initio} TDM surfaces for 1064~nm, 1550~nm, and 10~$\mu$m, respectively.  The
red shaded areas indicate the estimated error \cite{supplement}.  The
vertical black dotted line indicates a typical experimental laser intensity
of 10~kW~cm$^{-2}$.  We used an initial molecule density of $0.4 \cdot
10^{11}~$cm$^{-3}$, temperature of $500$~nK, and $p$-wave diatom-diatom
collision rate of $3 \cdot 10^{-11}$ cm$^{3}$s$^{-1}$.  The NaK lifetime,
$\tau_\mathrm{NaK}$, accounts for both laser excitation and
sticking-amplified three-body loss.}
\label{fig:tau_intens}
\end{figure}

Figure \ref{fig:tau_intens} shows the laser-excitation lifetime,
$\tau_\mathrm{laser}$, of the complexes (in blue, left-hand axis) and the
half-life of  the NaK molecules in the trap
$\tau_\mathrm{NaK}$ (in red, right-hand axis), as a function of the  laser
intensity.  The solid, dashed, and dash-dotted lines show results for
wavelengths of 1064~nm, 1550~nm, and 10~$\mu$m, respectively.  The lifetime
was calculated using an initial trap molecule density of $0.4 \cdot 10^{11}
$cm$^{-3}$, temperature of 500~nK \cite{park:2015}, and a diatom-diatom
$p$-wave collision rate of $3 \cdot 10^{-11}$~cm$^3$~s$^{-1}$ based on
multichannel quantum defect theory \cite{mayle:2013}.  We include both the
laser excitation loss mechanism and sticking-amplified three-body loss
with a rate of $1.1 \cdot 10^{-10}$~cm$^{-3}$s$^{-1}$, as
described in Ref.~\cite{christianen:2019b}.  The estimated lifetime of the
complex with respect to complex-molecule collisions $\tau_{3b}=1/[n(0)
k_{3b}]$, where $n(0)$ is the initial molecular density and $k_{3b}$ the
molecule-complex collision rate.
At 1064~nm and 1550~nm, in the typical range of experimental
intensities, we find laser excitation leads to effective two-body loss,
limited by the formation of collision complexes, and is insensitive to small
changes in wavelength and intensity.  For a 10~$\mu$m wavelength, however,
the lifetime of the molecules becomes orders of magnitude larger and
strongly intensity-dependent.  Changing the trapping laser wavelength to
10~$\mu$m emerges as a straightforward way to strongly reduce the losses.
However, even at this wavelength the laser excitation of collision complexes
is dominant over sticking-amplified three-body loss.

Possibilities for reducing trapping-laser-induced loss of bialkali molecules are the following:

1. Box potentials.  Rather than switching to very long wavelengths, a
promising avenue may be to use shorter wavelengths to realize repulsive
potential walls of blue-detuned trapping light \cite{davidson:95}, such that
molecules are trapped between them in the dark.  Such uniform box potentials
have been realized for ultracold atoms, for both Bose-Einstein condensates
\cite{gaunt:13} and Fermi gases \cite{mukherjee:17}.

2. Preventing molecular collisions. If molecular collisions are suppressed,
fewer collision complexes are formed that can be excited by the lasers.
Several ways to prevent molecular collisions have been proposed, such as
using optical lattices to confine the molecules \cite{miranda:11}, or
inducing repulsive interactions between colliding molecules using static
\cite{gonzalez:17}, or using microwave fields
\cite{karman:2018,lassabliere:2018}.

3. The trapping-laser excitation loss mechanism would be suppressed
completely if the molecules were confined by non-optical traps, such as a
magnetic trap.  This would require preparation of the molecules in a
low-field seeking Zeeman state, and may require nonzero electron spin.

We note that any additional effects not taken into account here---such as
an increased sticking time due to external fields or hyperfine coupling, or
excitations to higher excited states---can only increase the loss.  Such
effects could increase trapping-laser-induced loss at lower intensities,
while they do not affect the high-intensity limit where complex formation is
the rate-limiting step, see Fig.~\ref{fig:tau_intens}.

To summarize, we find that experimentally observed losses of nonreactive
ultracold molecules cannot be attributed to sticky collisions, but they can
be explained by electronic excitation of collision complexes by the trapping
laser.  The loss can be reduced by using significantly lower intensities and
longer wavelengths.  The loss can be prevented by using a magnetic instead
of optical trap, repulsive box potentials, or by preventing molecular
collisions.  The theory is illustrated for NaK+NaK collisions.  Other
bialkalis, such as NaRb and RbCs, have comparable electronic structure and
longer sticking times, such that we expect trapping-laser-induced loss to
also be the major cause of losses for these systems.

\section{Acknowledgements}

We thank Simon Cornish, Alan Jamison, Edvardas Narevicius, Kang-Kuen Ni,
Sebastian Will, and Zoe Yan for useful discussions.  M.Z. acknowledges
support from NSF, AFOSR, and the Gordon and Betty Moore Foundation through
grant GBMF5279.  T.K. is supported by NWO Rubicon grant 019.172EN.007 and an
NSF grant to ITAMP.

\bibliography{manuscript}

\end{document}